\documentclass[manuscript, screen]{acmart}

\setcopyright{none}
\renewcommand\footnotetextcopyrightpermission[1]{}

\makeatletter
\renewcommand\@mkauthorsaddresses{} 
\makeatother

\AtBeginDocument{%
  }

\setcopyright{acmlicensed}
\copyrightyear{2026}
\acmYear{2026}
\acmDOI{XXXXXXX.XXXXXXX}
\acmConference[LAK ’26]{the 16th International Learning Analytics \& Knowledge Conference}{April 27 – May 1, 2026}{Bergen, Norway}
\acmISBN{978-1-4503-XXXX-X/2018/06}

\usepackage{amsmath, amssymb, amsfonts}

\usepackage{algorithm}          
\usepackage[noend]{algpseudocode} 

\usepackage{placeins}

\begin{document}

\title{Feedback Effects on Cognitive Dynamics: Network-Based Insights from EEG Patterns and Behavioral Performance}


\author{Behdokht Kiafar}
\author{Mohammad Fahim Abrar}

\author{Roghayeh Leila Barmaki}







\renewcommand{\shortauthors}{Trovato et al.}

\begin{teaserfigure}
    \hspace{1.5cm}
    \includegraphics[width=0.85\textwidth]{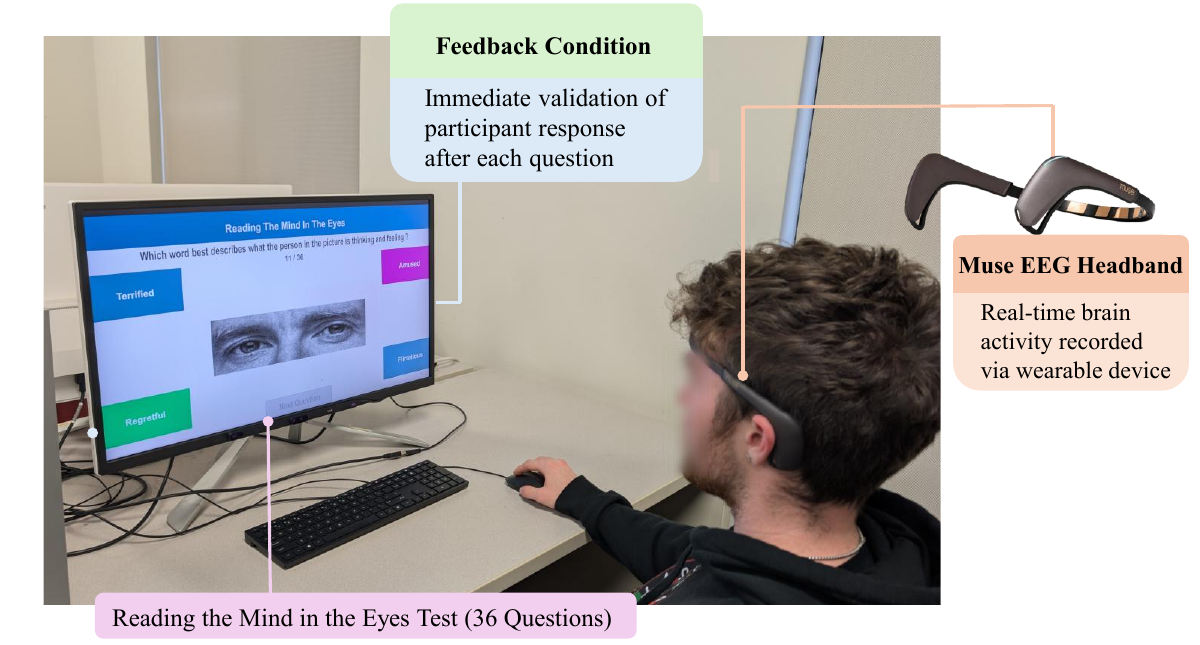}
    \caption{A participant completing the Reading the Mind in the Eyes Test while wearing a Muse EEG headband. In the Feedback condition,  participants received immediate indication of whether their response was correct or incorrect, whereas no such cue was provided in the No-Feedback condition.}
    \label{fig:Gaze}
\end{teaserfigure}

\begin{abstract}
  This study examines the impact of feedback on Electroencephalography (EEG) activity and performance during the Reading the Mind in the Eyes Test. In a within-subject design, eleven participants completed the test under Feedback and No-Feedback conditions. Using the principles of Epistemic Network Analysis (ENA) and Ordered Network Analysis (ONA), we extend these network-based models to explore the link between neural dynamics and task outcomes. ENA results showed that feedback is associated with stronger connections between higher frequency EEG bands (Beta and Gamma) and correct responses, while the absence of feedback activated lower frequency bands (Theta and Alpha). ONA further disclosed directional shifts toward higher frequency activity preceding correct answers in the Feedback condition, whereas the No-Feedback condition showed more self-connections in lower bands and a higher occurrence of wrong answers, suggesting less effective reasoning strategies without feedback. 
  Both ENA and ONA revealed statistically significant differences between conditions $(p = 0.01, Cohen’s d > 2)$.
  This study highlights the methodological benefits of integrating EEG with ENA and ONA for network analysis, capturing both temporal and relational dynamics, as well as the practical insight that feedback can foster more effective reasoning processes and improve task performance.
  {The code and setup to reproduce the experiments are publicly available \href{https://anonymous.4open.science/r/LAK-CC41/README.md}{\color{blue}here}}.
\end{abstract}




\maketitle

\section{Introduction}

Understanding how learners process information and respond to feedback is a central question in learning analytics. Feedback plays an important role in learning outcomes, and affects not only performance but also the cognitive strategies learners use \cite{gan2021teacher}. To capture these dynamics, researchers have turned to multimodal data sources that extend beyond traditional assessments, such as brain activity, gaze, or physiological signals \cite{mu2020multimodal}, which provide richer perspectives on learning in real time.
Electroencephalography (EEG) is one such modality that provides high temporal resolution for examining learners’ cognitive engagement \cite{executive2019ets}. However, EEG data are often challenging to interpret with conventional linear plots or frequency band summaries, limiting their use for educators and learning scientists. This gap underscores the need for analytic frameworks that can connect cognitive signals to learning processes in ways that are interpretable for learning analytics.

Methodological frameworks such as Epistemic Network Analysis (ENA) and Ordered Network Analysis (ONA) have shown strong potential for modeling learning processes by revealing relationships and temporal dynamics among elements of learner activity \cite{shaffer2016tutorial, tan2022epistemic}.
By integrating these frameworks with EEG, in this paper, we aim to bridge the gap between raw neural signals and interpretable representations of how feedback influences cognitive states and performance. This enables us to move beyond isolated spectral features toward dynamic representations and deeper insights into mechanisms of effective learning. Our contributions are as follows:

\begin{itemize}
  \item Introduction of two novel network methods: we present Neuro-Epistemic Network Analysis (NENA) and Neuro-Ordered Network Analysis (NONA), two complementary extensions of ENA and ONA, adapted to include EEG data.

  \item Application of the proposed frameworks:  we apply NENA and NONA to a pilot controlled study, using the Reading the Mind in the Eyes Test (RMET) \cite{baron2001reading}, and compare participant performance under two conditions of with and without immediate feedback.
  
  \item Empirical insights into feedback effects: we demonstrate how feedback shapes the structure and temporal dynamics of EEG connectivity, with interpretable evidence of its impact on neural activity.
  
  \item Open-source framework: we release our code and pipeline to support reproducibility in future EEG-based analytics studies.
\end{itemize}


\section{Related Work}

\subsection{Feedback in Learning Analytics}

Feedback has been recognized as a key component of the learning process, as it influences how learners engage and improve in both theory and practice. Vollmeyer and Rheinberg \cite{vollmeyer2005surprising} showed that even the anticipation of feedback encourages learners to adopt more systematic strategies, which in turn enhances knowledge acquisition and performance. Using a computer-simulated biology lab task, they compared learners who expected feedback with those who did not. Results showed that the expectation of feedback alone can trigger deeper processing and improve learning outcomes. Nicol and Macfarlane‐Dick \cite{nicol2006formative} further identified seven principles of feedback that promote self-regulation, which make students active agents in assessment rather than passive recipients in their learning. Extending this perspective, Pardo and Siemens \cite{pardo2014ethical} suggested that learning analytics must ensure that feedback is transparent and ethically grounded, and emphasized the importance of trust and accountability in its use. While prior work has established the value of feedback in shaping strategies and motivating learner engagement, understanding the neural processes underlying learning dynamics is also an important direction, which this preliminary study aims to address.

\subsection{EEG Applications in Education and Learning}
Researchers have increasingly explored how neurophysiological measures, such as EEG, can deepen our understanding of learning processes. For instance, Babiker et al. \cite{babiker2019eeg} demonstrated that EEG features, especially from Gamma and Delta bands, can be used to detect situational interest of students with high accuracy, even from a single frontal channel. In another study, Zander and Kothe \cite{zander2011towards} introduced passive brain–computer interfaces, which extend traditional active user input to infer cognitive and affective states in real time. In a related study, Wang et al. \cite{wang2025eeg} used EEG to examine how advanced digital tools impact student creativity and cognitive states during a design task. Students using intelligent systems showed higher creative performance and concentration than those using traditional software, with no significant difference in relaxation levels. These studies suggest that EEG provides valuable insights into learners’ engagement and cognitive strategies. However, most analyses remain static or descriptive. A promising next step is to apply frameworks that show how neural activity relates to learning strategies through temporal and relational patterns.

\subsection{Quantitative Ethnography and Network-Based Models}

Quantitative Ethnography (QE) provides a methodological framework to model connections in discourse, behavior, and context using network-based statistics \cite{shaffer2017quantitative}. As a core QE tool, ENA has been applied to examine collaborative problem-solving \cite{jiang2025using, zhang2023applied}. Fang et al. \cite{fang2024neural} combined Graph Neural Networks and ENA to analyze the sociocognitive nature of collaboration, with a focus on cognitive strategies in professional development. Zhou and Han \cite{zhou2025using} used ENA to study cognitive engagement in online collaborative learning, showing how different performance groups strategize and interact. Building on ENA, researchers have introduced ONA to capture temporal patterns by adding directionality to the model. For instance, directional ENA has been applied in online collaborative inquiry to examine how assessment and feedback influence critical thinking and group discourse \cite{ba2024anatomizing}. While prior studies have mainly combined ENA and ONA with behavioral and log-based data, little work has applied these frameworks to neurophysiological signals such as EEG. To our knowledge, this is one of the first efforts to integrate EEG into the QE framework, which extends its use to the analysis of cognitive states and learning processes.

\section{Materials and Methods}

We conducted a controlled user study to examine how feedback influences cognitive dynamics during the Reading the Mind in the Eyes Test. This involved preprocessing the EEG data, extracting spectral features, and analyzing the resulting signals using our proposed frameworks, Neuro-Epistemic Network Analysis and Neuro-Ordered Network Analysis, which extend ENA and ONA by incorporating neural signals into network models. Each step is described in the following sections.

\subsection{Participants and Task Design}

The study was reviewed and approved by the Institutional Review Board (IRB) at the authors’ institution, and informed consent was obtained from all volunteers. In total, 11 participants completed the RMET, a classic measure used in the theory of mind analysis. The within-subject design included two desktop-based conditions:
\begin{itemize}
    \item Feedback: participants received immediate indication of whether their answer was correct or not.
    \item No-Feedback: participants did not receive any information about the accuracy of their responses.
\end{itemize}
Each participant completed 36 multiple-choice RMET questions, with 18 assigned to each condition in a counterbalanced order. Each question presents a photograph of a person’s eyes, and participants select the word that best describes what the person is thinking or feeling from four options. The average study duration was $14.5$ minutes per participant $(SD = 1.2)$, including briefing, EEG headband setup, task execution, and debriefing.


\subsection{EEG Data Collection and Preprocessing}

\begin{figure*}
\centering
\includegraphics[width=0.3\textwidth]{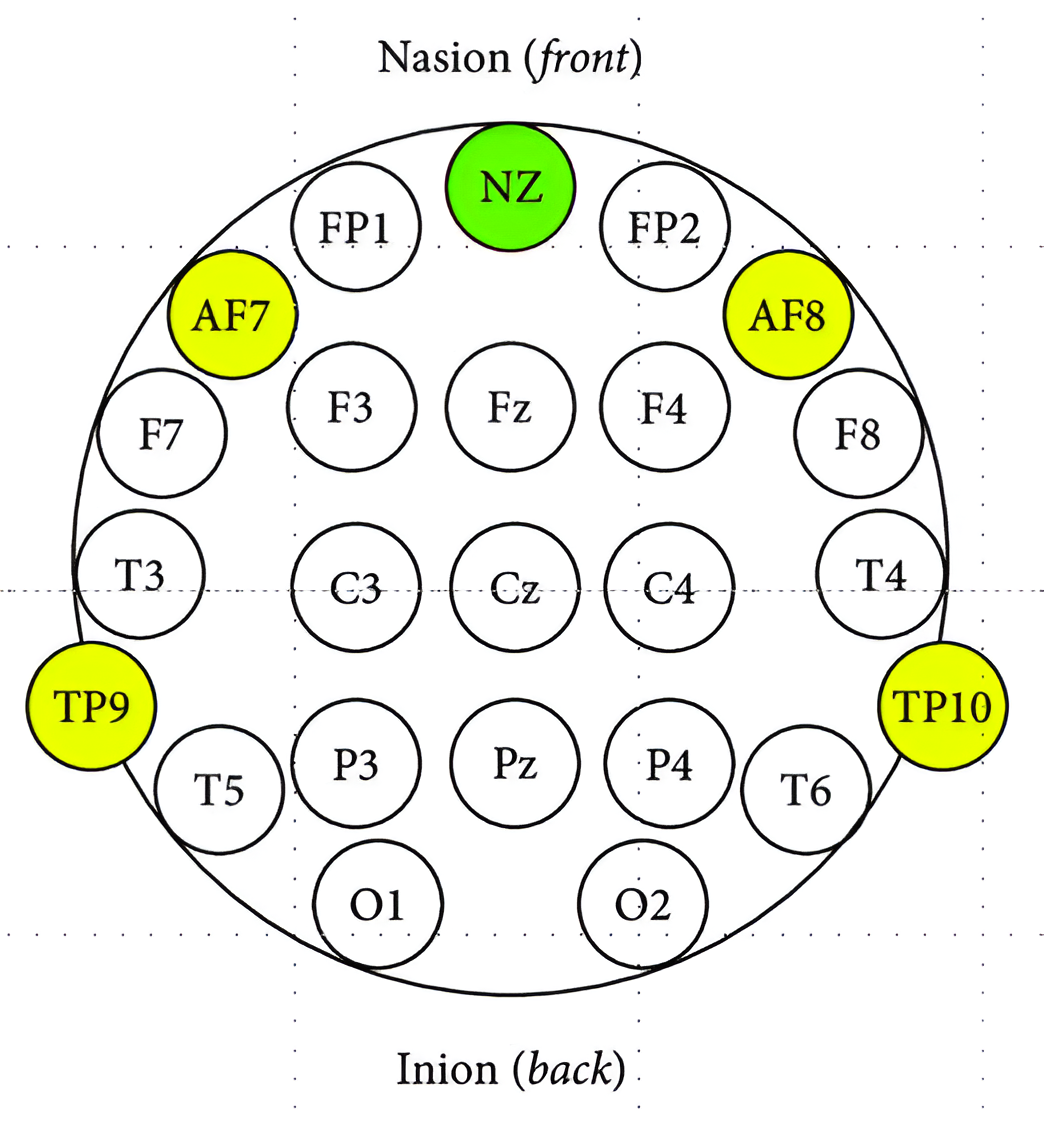}
\caption{Electrode positions of the Muse EEG headband channels used for data collection, according to the standard 10–20 system \cite{jasper1958ten}.}
\label{fig:Muse}
\end{figure*}

We recorded EEG using the Muse headband (InteraXon Inc., Toronto, Canada), a wearable device equipped with four electrodes (TP9, TP10, AF7, AF8) at a sampling rate of 256 Hz. Figure \ref{fig:Muse} shows the configuration of the Muse headband electrodes.
We first re-referenced the raw data using the average reference method \cite{yao2017surface} to minimize noise resulting from the location of the original reference electrode. Next, we applied a notch filter to remove interference from the 60 Hz alternating current power grid. We then used a band-pass filter to isolate data between 1–50 Hz, which included the $\delta$ (1–3 Hz), $\theta$ (4–7 Hz), $\alpha$ (8–13 Hz), $\beta$ (14–30 Hz), and $\gamma$ (31–49 Hz) frequency bands. We divided the data into 1-second epochs, each containing 256 samples. Finally, we applied Independent Component Analysis (ICA) to remove artifacts caused by eye and muscle activities \cite{winkler2011automatic}.

\subsection{Feature Extraction}
After cleaning and segmenting the EEG data, the next step is to extract features suitable for input to our network models. Since ENA and ONA rely on standardized binary input, we transformed raw EEG into frequency bands that are indicative of underlying cognitive states. We implemented this transformation by estimating the Power Spectral Density (PSD) with the Welch method \cite{zhang2023applied}. The process begins by dividing each epoch signal into three overlapping segments \cite{proakis2001digital}. We applied a Hamming window to each segment before performing the Fast Fourier Transform. Then we calculated a periodogram for each segment and averaged the results to obtain the spectral estimates. 
To achieve binary formats, we integrated the PSD over each band and calculated its contribution as a percentage of the total power across the full frequency spectrum.
We applied a Signal-to-Noise Ratio (SNR) \cite{cerutti2011advanced} threshold to determine the significance of each band’s contribution to the overall EEG activity within a segment. Based on this threshold, we encoded the presence or absence of each band in binary form per epoch, and finally, used majority voting across channels to consolidate frequency activity into a single representation.

\subsection{Network-Based Methods}

\subsubsection{Neuro-Epistemic Network Analysis (NENA)}

In this method, we extend ENA, a relatively new analytical approach originally developed to model co-occurrence patterns in discourse data \cite{shaffer2016tutorial}.
ENA represents elements of interest as nodes and captures their co-occurrence as edges, where node size indicates frequency of occurrence and edge thickness reflects the strength of association.
NENA adapts this framework to neurophysiological data. To prepare EEG data for analysis, we constructed binary vectors for each epoch that encoded the presence of frequency bands along with the correctness of responses. These vectors were accumulated into symmetric co-occurrence matrices, normalized by the number of epochs per participant, and then projected into a lower-dimensional space using Singular Value Decomposition (SVD) \cite{bowman2021mathematical}.
The resulting networks are represented as graphs where EEG frequency bands and response accuracy (Correct, Incorrect) appear as nodes, and undirected edges capture their co-occurrence.
This approach provides interpretable visualizations of how neural activity aligns with task performance and offers insight into the cognitive processes engaged during the task. It also reveals how the strength and distribution of these connections vary across experimental conditions.

\subsubsection{Neuro-Ordered Network Analysis (NONA)}

While NENA focuses on co-occurrence patterns, in NONA, we extend ONA to model how connections change over time. ONA is a dynamic network approach designed to capture how connections among elements evolve by incorporating temporal directionality \cite{tan2022ordered}. This framework is particularly useful in contexts where the sequence of events is hypothesized to be meaningful, such as in feedback scenarios during learning, where the ordered nature of events can influence subsequent cognitive states.
ONA visualizations use a broadcast model in which each edge is shown as a pair of triangles where the apex indicates the origin (ground) and the base points to the destination (response). Dark chevrons inside the triangles mark the direction of flow. Node size shows how often a node appears as a response, and the saturation of its inner circle denotes the number of self-connections, representing repeated activations of the same node across time.
In NONA, we adapt this framework to neurophysiological data by modeling directed co-occurrences between EEG frequency bands and performance outcomes across consecutive time windows. This approach enables the investigation of how feedback influences the progression of cognitive states and demonstrates how earlier neural activity may impact later task performance.
The pipeline for implementing NENA and NONA with EEG data is presented in Algorithm \ref{alg:unified-nena-nona}, and the source code for preparing the data and running the algorithm is accessible \href{https://anonymous.4open.science/r/LAK-CC41/README.md}{\color{blue}here}.


\FloatBarrier
\begin{algorithm}
\small
\caption \centering{Pipeline for applying NENA and NONA}
\label{alg:unified-nena-nona}

\vspace{0.1 cm}
\textbf{Inputs:} EEG channels $\{TP9, TP10, AF7, AF8\}$, $f_s\!=\!256$ Hz; Bands $\{\delta,\theta,\alpha,\beta,\gamma\}$; Per-item labels: $\text{Response} \in \{\text{Correct}, \text{Incorrect}\}$, $\text{Condition} \in \{\text{Feedback},\text{No-Feedback}\}$; NONA window $L$; Units $U$ (participants); {Codes} $C=\{\delta,\theta,\alpha,\beta,\gamma,$ Correct, Incorrect$\}$

\begin{algorithmic}[1]

\Statex

\For{each unit $u \in U$}
  \State \textbf{Preprocess EEG}: average reference; 60 Hz notch; 1–50 Hz band-pass; 1 s epochs; ICA artifact removal
  \For{each epoch $e$ and channel}
    \State \textbf{Welch PSD} $\rightarrow$ band share $=$ band\_power / total\_power; apply SNR threshold $\rightarrow$ binary presence per band
  \EndFor
  \State \textbf{Majority vote} across channels per band; append one-hot \{Correct $\vert$ Incorrect\} to form binary code vector $x_e$
  \Statex

  \State \textbf{NENA (symmetric co-occurrence)}
  \State initialize $A_{\text{sym}} \leftarrow \mathbf{0}_{|C| \times |C|}$
  \For{each epoch $e$}
    \State $S \leftarrow \{ \text{active codes in } x_e \}$
    \For{each unordered pair $(i,j)$ in $S \times S$, $i \neq j$} $A_{\text{sym}}[i,j] \mathrel{+}= 1$ \EndFor
  \EndFor
  \State $v_{\text{sym}} \leftarrow \text{UpperTriangle}(A_{\text{sym}})$
  \Statex

  \State \textbf{NONA (directed co-occurrence)}
  \State initialize $A_{\text{dir}} \leftarrow \mathbf{0}_{|C| \times |C|}$
  \For{each epoch $t$}
    \State $G \leftarrow$ active codes in $[t-(L-1), \ldots, t-1]$ \Comment{ground}
    \State $R \leftarrow$ active codes in $t$ \Comment{response}
    \For{each $(g,r) \in G \times R$} $A_{\text{dir}}[g,r] \mathrel{+}= 1$ \EndFor
  \EndFor
  \State $v_{\text{dir}} \leftarrow \text{Flatten}(A_{\text{dir}})$
\EndFor
\Statex

\State \textbf{Normalization}
\For{each unit $u$}
  \State scale $v_{\text{sym}}[u]$ by number of epochs for $u$ (or sum of $A_{\text{sym}}[u]$ entries)
  \State scale $v_{\text{dir}}[u]$ by number of window updates for $u$ (or sum of $A_{\text{dir}}[u]$ entries)
\EndFor

\Statex
\State \textbf{Dimensionality Reduction and Projection}
\State stack $V_{\text{sym}} \leftarrow \{v_{\text{sym}}[u]\}_{u \in U}$, $V_{\text{dir}} \leftarrow \{v_{\text{dir}}[u]\}_{u \in U}$
\State compute separate 2D Singular Value Decomposition projections for $V_{\text{sym}}$ and $V_{\text{dir}}$
\State fix node positions using ENA optimization for consistent comparisons

\end{algorithmic}
\end{algorithm}

\section{Results}
\subsection{Neuro-Epistemic Network Analysis}

\begin{figure*} [h]
    \centering     \includegraphics[width=1\textwidth]{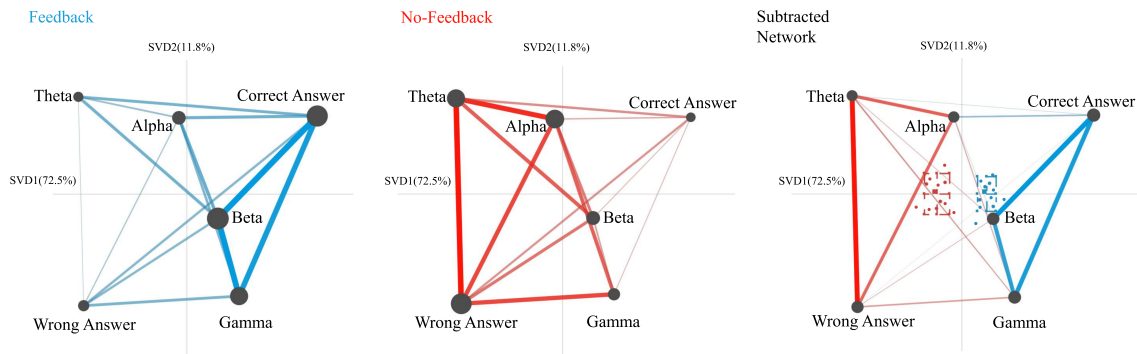}
    \caption{NENA representations of EEG activity and task performance during the RMET: Feedback (left), No-Feedback (middle), and the subtracted network (right). In the model, the size of each node corresponds to the frequency with which the node appears, while the thickness of the lines between nodes indicates the strength of the connections. In general, participants demonstrated denser connections with higher frequency bands in the Feedback condition.
 }

    \label{fig:NENA Networks}
    
\end{figure*}

After applying NENA and fixing node positions, we compared networks in a two-dimensional space by examining node connections and their strengths. Figure \ref{fig:NENA Networks} presents the networks for both conditions: the blue network corresponds to the Feedback condition, and the red network corresponds to the No-Feedback condition.
The rightmost network depicts the subtracted graph, which provides pairwise differences generated by subtracting the weights of each connection in one network from the corresponding connections in the other network. Centroids appear as squares, and dotted lines indicate 95\% confidence intervals.
The first dimension (SVD1) explains 72.5\% of the variance in the structure of connections, and the second dimension (SVD2) explains 11.8\%.
In the feedback (blue) network, we observed denser connections among higher frequency EEG bands, particularly Beta and Gamma, which also showed strong co-occurrence with correct answers.
In contrast, in the No-Feedback (red) network, we observed stronger links among Theta and Alpha bands, which were more frequently tied to wrong answers.
The subtracted network further highlights these contrasts; edges among Beta, Gamma, and correct responses are more prominent in the Feedback condition, while edges involving Theta, Alpha, and incorrect responses dominate in the No-Feedback condition.
Along the \textit{X-axis}, a two-sample t-test assuming unequal variances, showed a statistically significant difference at the $\alpha = 0.05$ level between Feedback condition ($mean = 0.41, SD = 0.10, N = 11$) and No-Feedback condition ($mean = -0.41, SD = 0.13, N = 11$); $p = 0.01$, Cohen’s d = $2.06$. No significant difference was found along the \textit{Y-axis}.

\subsection{Neuro-Ordered Network Analysis}

\begin{figure*} [h]
    \centering     \includegraphics[width=1\textwidth]{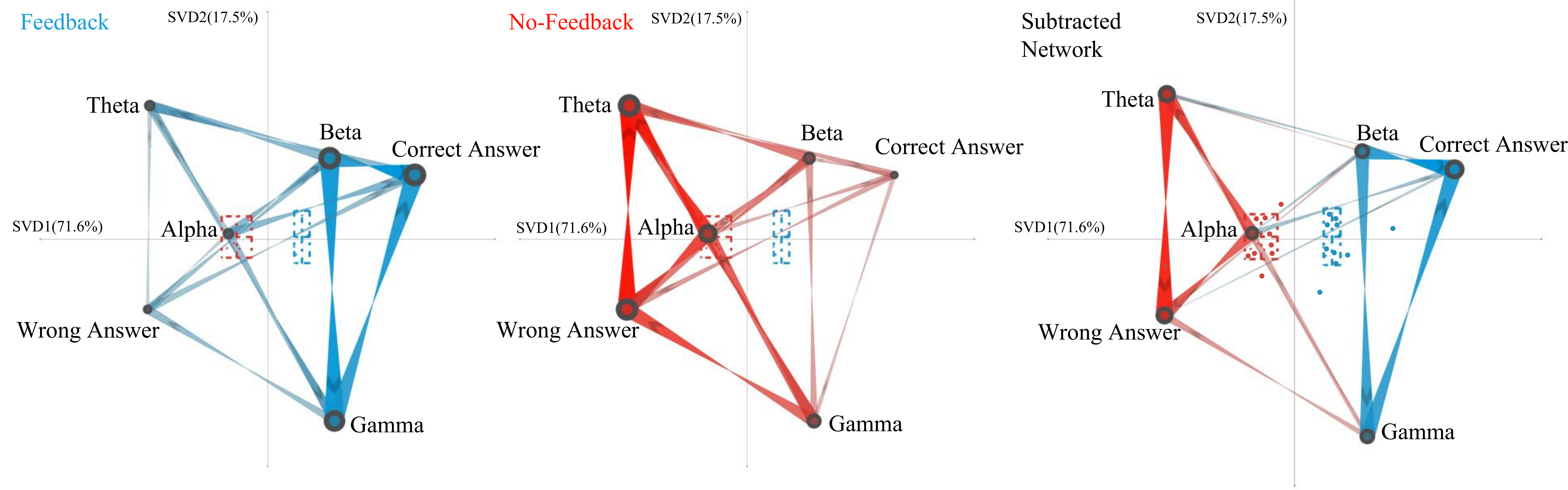}
    \caption{NONA representations of EEG activity and task performance during the RMET under two conditions: Feedback (left), No-Feedback (middle), and the subtracted network (right). The directed edges (indicated by chevrons) illustrate the temporal sequence of co-occurrences.
 }

    \label{fig:NONA Networks}
    
\end{figure*}

Figure \ref{fig:NONA Networks} shows NONA networks for both conditions, with blue representing Feedback condition, red indicating No-Feedback condition, and the right panel showing their differences.
In these visualizations, directed edges (chevrons) denote the temporal flow of co-occurrences, node size indicates response frequency, and the inner circle within each node represents self-connections, which indicate that the same construct continues to reappear in consecutive windows.
The amount of variance explained by the variables represented in axes SVD1 and SVD2 are 71.6\% and 17.5\% respectively.
In the Feedback condition, we observed forward-directed edges leading to Gamma, particularly on correct trials. By contrast, in the No-Feedback condition, Theta and Alpha showed more self-connections and links to other constructs, especially to wrong answers.
A two-sample t-test was conducted to evaluate whether there were significant differences between the means of each condition.
Along the \textit{X-axis}, the t-test assuming unequal variances, indicated a statistically significant difference at the $\alpha = 0.05$ level between Feedback condition ($mean = 0.41, SD = 0.07, N = 11$) and No-Feedback condition ($mean = -0.41, SD = 0.12, N = 11$); $p = 0.01$, Cohen’s d = $2.18$.
Along the \textit{Y-axis}, no significant difference was found.

\section{Discussion}

In the NENA results, stronger connections are observed between higher-frequency EEG bands, particularly Beta and Gamma, when participants answered questions correctly with feedback. This pattern suggests a heightened engagement and cognitive activity, likely stimulated by the immediate feedback. Conversely, in the absence of feedback, Theta and Alpha frequency bands showed more significant connections. This pattern may indicate a different cognitive approach where participants are possibly more reliant on memory recall or internal processing rather than real-time problem-solving and adjustment based on external cues.
This difference underscores how the absence of feedback might lead participants to engage less in inference, interpretation, or the integration of information, as previously noted in literature \cite{barth2017evaluating, hall2020effects}.

In NONA, the Feedback condition shows more forward-directed links toward Gamma on correct trials. These stronger forward-directed connections suggest that feedback may have promoted a progression of cognitive states associated with active reasoning, commonly attributed to higher-frequency EEG activity \cite{chikhi2022eeg, rosen2017right}.
In contrast, the No-Feedback condition shows more prominent self-connections for Theta and Alpha, along with noticeable links between these bands and other constructs. The subtracted network highlights these differences and suggests that feedback on correct answers encourages participants to provide more correct responses and activates higher-frequency bands, as reflected in the increased self-connections. Similar findings are reported in previous work \cite{acklin2012effects}, where positive feedback is shown to be more closely associated with high-level cognitive processes.
The chevron on the edges between higher frequency bands and correct answers in the Feedback condition suggests that activation of these bands may contribute to a higher number of correct responses. Additionally, the chevron between incorrect and correct answers indicates that participants are more likely to respond correctly after receiving feedback on incorrect answers compared to the other condition.
Practically, this suggests that timely feedback can support more efficient problem-solving in social cognition tasks.

Taken together, our work shows how network-based approaches can make neural dynamics more interpretable by linking EEG patterns to feedback-related cognitive processes. These models not only affirm established findings on learning patterns but also reveal the neural mechanisms behind them. As such, NENA and NONA can be powerful tools in learning analytics for visualizing the neural underpinnings of cognitive processes by capturing the neural basis through EEG signals.
\paragraph{Closing the Interpretive Loop}
We closed the interpretive loop between raw neural signals, network models, and theoretical constructs of learning, which is a key goal of Quantitative Ethnography \cite{arastoopour2022foundations}. Our analyses showed how feedback strengthened co-occurrences between higher frequency activity and correct responses, and how temporal progressions in neural states aligned with more effective reasoning strategies \cite{dickey2022widespread, baker2025time}. This connection ensures that the results are theoretically grounded in the learning sciences.

\paragraph{Limitations and Future Directions} Like any preliminary study, this work has some limitations.
First, EEG signals are inherently prone to noise and non-neural artifacts, which may introduce variability into the analysis.
Additionally, we investigated the immediate neural dynamics of feedback provision during the intervention.
In the future, we plan to address issues related to common sources of noise in EEG data \cite{delorme2023eeg} by incorporating visual representations of uncertainty into our framework to enhance its interpretability. We also intend to extend the scope of our analysis by examining knowledge retention at follow-up intervals (e.g., 3 or 7 days post-intervention), which can provide a richer understanding of retention and learning in our experiments.
Another important direction for future work is the integration of machine learning techniques \cite{ mohammadagha2025hyperparameter, mohammadagha2025hybridization, mohammadagha2025evaluating} with the proposed network-based frameworks. Such hybrid approaches may also support more adaptive or scalable analyses across larger datasets \cite{mohammadagha2025machine, mohammadagha2025cross, mohammadagha2025hybrid, mohammadagha2025comparative} and diverse learning contexts.

\section{Conclusion}

This paper introduced two network-based methods, Neuro-Epistemic Network Analysis and Neuro-Ordered Network Analysis, to connect EEG dynamics with task performance under different feedback conditions during the RMET, a classic test in the theory of mind assessment. By extending ENA and ONA to brain signals, these approaches move from raw spectral power to interpretable network maps that foreground how frequency bands relate to correct and incorrect responses and how these relations evolve over time.
Across conditions, the NENA analyses found that feedback is associated with stronger alignment between higher frequency activity and correct responses, whereas the absence of feedback is linked to lower frequency activity and weaker connections to correct answers. NONA additionally showed directional shifts toward higher frequency bands preceding correct answers when feedback was present. The subtracted networks and comparative analyses also provided clear visual and statistical evidence that feedback provision changes both the structure and the temporal flow of neuro-behavioral connections.
Overall, our models offer an intuitive framework that connects EEG patterns with task outcomes by converting raw signals into network representations of cognition, which reveal feedback-based processes that traditional methods cannot capture. The frameworks developed here can be extended beyond the RMET to other learning and assessment contexts, where the integration of neural and behavioral data may guide the design of adaptive environments.


\begin{acks}
To Robert, for the bagels and explaining CMYK and color spaces.
\end{acks}

\bibliographystyle{ACM-Reference-Format}
\bibliography{sample-base}

@article{shaffer2016tutorial,
  title={A tutorial on epistemic network analysis: Analyzing the structure of connections in cognitive, social, and interaction data},
  author={Shaffer, David Williamson and Collier, Wesley and Ruis, Andrew R},
  journal={Journal of learning analytics},
  volume={3},
  number={3},
  pages={9--45},
  year={2016}
}

@inproceedings{tan2022ordered,
  title={Ordered network analysis},
  author={Tan, Yuanru and Ruis, Andrew R and Marquart, Cody and Cai, Zhiqiang and Knowles, Mariah A and Shaffer, David Williamson},
  booktitle={International Conference on Quantitative Ethnography},
  pages={101--116},
  year={2022},
  organization={Springer}
}

@article{baron2001reading,
  title={The “Reading the Mind in the Eyes” test revised version: A study with normal adults, and adults with Asperger syndrome or high-functioning autism},
  author={Baron-Cohen, Simon and Wheelwright, Sally and Hill, Jacqueline and Raste, Yogini and Plumb, Ian},
  journal={Journal of child psychology and psychiatry},
  volume={42},
  number={2},
  pages={241--251},
  year={2001},
  publisher={Wiley Online Library}
}

@article{yao2017surface,
  title={Is the surface potential integral of a dipole in a volume conductor always zero? A cloud over the average reference of EEG and ERP},
  author={Yao, Dezhong},
  journal={Brain Topography},
  volume={30},
  number={2},
  pages={161--171},
  year={2017},
  publisher={Springer}
}

@article{winkler2011automatic,
  title={Automatic classification of artifactual ICA-components for artifact removal in EEG signals},
  author={Winkler, Irene and Haufe, Stefan and Tangermann, Michael},
  journal={Behavioral and brain functions},
  volume={7},
  pages={1--15},
  year={2011},
  publisher={Springer}
}

@book{proakis2001digital,
  title={Digital signal processing: principles algorithms and applications},
  author={Proakis, John G},
  year={2001},
  publisher={Pearson Education India}
}

@book{cerutti2011advanced,
  title={Advanced methods of biomedical signal processing},
  author={Cerutti, Sergio and Marchesi, Carlo},
  year={2011},
  publisher={John Wiley \& Sons}
}

@inproceedings{bowman2021mathematical,
  title={The mathematical foundations of epistemic network analysis},
  author={Bowman, Dale and Swiecki, Zachari and Cai, Zhiqiang and Wang, Yeyu and Eagan, Brendan and Linderoth, Jeff and Shaffer, David Williamson},
  booktitle={Advances in Quantitative Ethnography: Second International Conference, ICQE 2020, Malibu, CA, USA, February 1-3, 2021, Proceedings 2},
  pages={91--105},
  year={2021},
  organization={Springer}
}

@article{delorme2023eeg,
  title={EEG is better left alone},
  author={Delorme, Arnaud},
  journal={Scientific reports},
  volume={13},
  number={1},
  pages={2372},
  year={2023},
  publisher={Nature Publishing Group UK London}
}

@article{barth2017evaluating,
  title={Evaluating the impact of a multistrategy inference intervention for middle-grade struggling readers},
  author={Barth, Amy E and Elleman, Amy},
  journal={Language, speech, and hearing services in schools},
  volume={48},
  number={1},
  pages={31--41},
  year={2017},
  publisher={American Speech-Language-Hearing Association}
}

@article{hall2020effects,
  title={The effects of inference instruction on the reading comprehension of English learners with reading comprehension difficulties},
  author={Hall, Colby and Vaughn, Sharon and Barnes, Marcia A and Stewart, Alicia A and Austin, Christy R and Roberts, Greg},
  journal={Remedial and Special Education},
  volume={41},
  number={5},
  pages={259--270},
  year={2020},
  publisher={Sage Publications Sage CA: Los Angeles, CA}
}

@inproceedings{tan2022epistemic,
  title={Epistemic network analysis visualization},
  author={Tan, Yuanru and Hinojosa, Cesar and Marquart, Cody and Ruis, Andrew R and Shaffer, David Williamson},
  booktitle={Advances in Quantitative Ethnography: Third International Conference, ICQE 2021, Virtual Event, November 6--11, 2021, Proceedings 3},
  pages={129--143},
  year={2022},
  organization={Springer}
}

@article{chikhi2022eeg,
  title={EEG power spectral measures of cognitive workload: A meta-analysis},
  author={Chikhi, Samy and Matton, Nadine and Blanchet, Sophie},
  journal={Psychophysiology},
  volume={59},
  number={6},
  pages={e14009},
  year={2022},
  publisher={Wiley Online Library}
}

@article{rosen2017right,
  title={Right frontal gamma and beta band enhancement while solving a spatial puzzle with insight},
  author={Rosen, A and Reiner, M},
  journal={International Journal of Psychophysiology},
  volume={122},
  pages={50--55},
  year={2017},
  publisher={Elsevier}
}

@phdthesis{acklin2012effects,
  title={The effects of feedback on working memory capacity},
  author={Acklin, William Thomas},
  year={2012},
  school={University of Georgia}
}

@article{zhang2023applied,
  title={The applied principles of EEG analysis methods in neuroscience and clinical neurology},
  author={Zhang, Hao and Zhou, Qing-Qi and Chen, He and Hu, Xiao-Qing and Li, Wei-Guang and Bai, Yang and Han, Jun-Xia and Wang, Yao and Liang, Zhen-Hu and Chen, Dan and others},
  journal={Military Medical Research},
  volume={10},
  number={1},
  pages={67},
  year={2023},
  publisher={Springer}
}

@article{nicol2006formative,
  title={Formative assessment and self-regulated learning: A model and seven principles of good feedback practice},
  author={Nicol, David J and Macfarlane-Dick, Debra},
  journal={Studies in higher education},
  volume={31},
  number={2},
  pages={199--218},
  year={2006},
  publisher={Taylor \& Francis}
}

@article{pardo2014ethical,
  title={Ethical and privacy principles for learning analytics},
  author={Pardo, Abelardo and Siemens, George},
  journal={British journal of educational technology},
  volume={45},
  number={3},
  pages={438--450},
  year={2014},
  publisher={Wiley Online Library}
}

@article{vollmeyer2005surprising,
  title={A surprising effect of feedback on learning},
  author={Vollmeyer, Regina and Rheinberg, Falko},
  journal={Learning and instruction},
  volume={15},
  number={6},
  pages={589--602},
  year={2005},
  publisher={Elsevier}
}

@article{zander2011towards,
  title={Towards passive brain--computer interfaces: applying brain--computer interface technology to human--machine systems in general},
  author={Zander, Thorsten O and Kothe, Christian},
  journal={Journal of neural engineering},
  volume={8},
  number={2},
  pages={025005},
  year={2011},
  publisher={IOP Publishing}
}

@article{babiker2019eeg,
  title={EEG in classroom: EMD features to detect situational interest of students during learning},
  author={Babiker, Areej and Faye, Ibrahima and Mumtaz, Wajid and Malik, Aamir Saeed and Sato, Hiroki},
  journal={Multimedia Tools and Applications},
  volume={78},
  number={12},
  pages={16261--16281},
  year={2019},
  publisher={Springer}
}

@article{gan2021teacher,
  title={Teacher feedback practices, student feedback motivation, and feedback behavior: how are they associated with learning outcomes?},
  author={Gan, Zhengdong and An, Zhujun and Liu, Fulan},
  journal={Frontiers in psychology},
  volume={12},
  pages={697045},
  year={2021},
  publisher={Frontiers Media SA}
}

@article{mu2020multimodal,
  title={Multimodal data fusion in learning analytics: A systematic review},
  author={Mu, Su and Cui, Meng and Huang, Xiaodi},
  journal={Sensors},
  volume={20},
  number={23},
  pages={6856},
  year={2020},
  publisher={MDPI}
}

@article{executive2019ets,
  title={ETS Research Report Series},
  author={EXECUTIVE, EIGNOR},
  year={2019}
}

@article{wang2025eeg,
  title={EEG assessment of artificial intelligence-generated content impact on student creative performance and neurophysiological states in product design},
  author={Wang, Shuxin and Tao, Xin and Ma, Hongbo and Li, Fanglian and Wu, Chuanqi},
  journal={Frontiers in Psychology},
  volume={16},
  pages={1508383},
  year={2025},
  publisher={Frontiers Media SA}
}

@inproceedings{jiang2025using,
  title={Using Epistemic Network Analysis and Sequential Pattern Mining to Explore the Impacts of Human Facilitation on Collaborative Mathematical Problem Solving},
  author={Jiang, Yang and Graf, Edith Aurora and Todd, Jessica Andrews},
  booktitle={Proceedings of the 18th International Conference on Computer-Supported Collaborative Learning-CSCL 2025, pp. 3-11},
  year={2025},
  organization={International Society of the Learning Sciences}
}

@inproceedings{fang2024neural,
  title={Neural epistemic network analysis: Combining graph neural networks and epistemic network analysis to model collaborative processes},
  author={Fang, Zheng and Wang, Weiqing and Chen, Guanliang and Swiecki, Zachari},
  booktitle={Proceedings of the 14th Learning Analytics and Knowledge Conference},
  pages={157--166},
  year={2024}
}

@article{zhou2025using,
  title={Using epistemic network analysis to examine the cognitive engagement process in online collaborative learning},
  author={Zhou, Jin and Han, Xuejing},
  journal={Acta Psychologica},
  volume={253},
  pages={104737},
  year={2025},
  publisher={Elsevier}
}

@article{ba2024anatomizing,
  title={Anatomizing online collaborative inquiry using directional epistemic network analysis and trajectory tracking},
  author={Ba, Shen and Hu, Xiao and Stein, David and Liu, Qingtang},
  journal={British Journal of Educational Technology},
  volume={55},
  number={5},
  pages={2173--2191},
  year={2024},
  publisher={Wiley Online Library}
}

@book{shaffer2017quantitative,
  title={Quantitative ethnography},
  author={Shaffer, David Williamson},
  year={2017},
  publisher={Lulu. com}
}

@article{dickey2022widespread,
  title={Widespread ripples synchronize human cortical activity during sleep, waking, and memory recall},
  author={Dickey, Charles W and Verzhbinsky, Ilya A and Jiang, Xi and Rosen, Burke Q and Kajfez, Sophie and Stedelin, Brittany and Shih, Jerry J and Ben-Haim, Sharona and Raslan, Ahmed M and Eskandar, Emad N and others},
  journal={Proceedings of the National Academy of Sciences},
  volume={119},
  number={28},
  pages={e2107797119},
  year={2022},
  publisher={National Academy of Sciences}
}

@article{baker2025time,
  title={Time-domain brain: Temporal mechanisms for brain functions using time-delay nets, holographic processes, radio communications, and emergent oscillatory sequences},
  author={Baker, Janet M and Cariani, Peter},
  journal={Frontiers in Computational Neuroscience},
  volume={19},
  pages={1540532},
  year={2025},
  publisher={Frontiers Media SA}
}

@inproceedings{arastoopour2022foundations,
  title={The foundations and fundamentals of quantitative ethnography},
  author={Arastoopour Irgens, Golnaz and Eagan, Brendan},
  booktitle={International Conference on Quantitative Ethnography},
  pages={3--16},
  year={2022},
  organization={Springer}
}

@article{jasper1958ten,
  title={Ten-twenty electrode system of the international federation},
  author={Jasper, Herbert H},
  journal={Electroencephalogr Clin Neurophysiol},
  volume={10},
  pages={371--375},
  year={1958}
}

@article{mohammadagha2025machine,
  title={Machine learning models for reinforced concrete pipes condition prediction: The state-of-the-art using artificial neural networks and multiple linear regression in a Wisconsin case study},
  author={Mohammadagha, Mohsen and Najafi, Mohammad and Kaushal, Vinayak and Jibreen, Ahmad Mahmoud Ahmad},
  journal={arXiv preprint arXiv:2502.00363},
  year={2025}
}

@article{mohammadagha2025hyperparameter,
  title={Hyperparameter Optimization Strategies for Tree-Based Machine Learning Models Prediction: A Comparative Study of AdaBoost, Decision Trees, and Random Forest},
  author={Mohammadagha, Mohsen},
  journal={Decision Trees, and Random Forest (April 11, 2025)},
  year={2025}
}

@article{mohammadagha2025hybridization,
  title={Hybridization and Optimization Modeling, Analysis, and Comparative Study of Sorting Algorithms: Adaptive Techniques, Parallelization, for Mergesort, Heapsort, Quicksort, Insertion Sort, Selection Sort, and Bubble Sort},
  author={Mohammadagha, Mohsen},
  year={2025},
  publisher={Engineering Archive}
}

@article{mohammadagha2025evaluating,
  title={Evaluating Machine Learning Performance Using Python for Neural Network Models in Urban Transportation in New York City Case Study},
  author={Mohammadagha, Mohsen and Asadi, Saeed and Naeini, Hajar Kazemi},
  year={2025}
}

@article{mohammadagha2025cross,
  title={Cross-Domain Applications of Machine Learning: A Comparative Case Study from Iris Classification to Infrastructure Assessment},
  author={Mohammadagha, Mohsen and Najafi, Mohammad and Kaushal, Vinayak and Jibreen, Ahmad},
  journal={Computer and Decision Making: An International Journal},
  volume={2},
  pages={742--765},
  year={2025}
}

@article{mohammadagha2025hybrid,
  title={Hybrid machine learning meta-model for the condition assessment of urban underground pipes},
  author={Mohammadagha, Mohsen and Najafi, Mohammad and Kaushal, Vinayak and Jibreen, Ahmad},
  journal={Infrastructures},
  volume={10},
  number={11},
  pages={282},
  year={2025},
  publisher={MDPI}
}

@article{mohammadagha2025comparative,
  title={Comparative Deep Learning Analysis of Regularization Techniques on Generalization in Baseline CNNs and ResNet Architectures for Machine Learning-Based Image Classification},
  author={Mohammadagha, Mohsen and Bigdeli, Farbod and Sharifi, Shayan and Deldadehasl, Maryam and Ataei, Saeid},
  year={2025},
  publisher={Preprints}
}


\end{document}